\documentclass[aps,prb,twocolumn,superscriptaddress,showpacs]{revtex4}
\usepackage{graphicx}

\newcommand{\be}{\begin{equation}}
\newcommand{\ee}{\end{equation}}
\renewcommand\vec[1]{\mbox{\boldmath $#1$}}
\def\p{^{\,\prime}}

\newcounter{appendixc}
\renewcommand{\appendix}[1]{\vspace*{0.6cm}
\refstepcounter{appendixc} \setcounter{table}{0}
\setcounter{equation}{0}
\renewcommand{\thetable}{\Alph{appendixc}\arabic{table}}
\renewcommand{\theappendixc}{\Alph{appendixc}}
\renewcommand{\theequation}{\Alph{appendixc}\arabic{equation}}
\noindent{\bf Appendix \theappendixc. #1}\par\vspace*{0.4cm}}

\begin{document}

\title{Electron-phonon interaction in quantum-dot/quantum-well semiconductor heterostructures}
\author{F. Comas}
\affiliation{Departamento de F\'{\i}sica Te\'{o}rica, Universidad
de la Habana, Vedado 10400, Havana, Cuba}
\affiliation{Departamento de F\'{\i}sica, Universidade Federal de
S\~ao Carlos, 13565-905 S\~ao Carlos SP, Brazil}
\author{Nelson Studart}
\affiliation{Departamento de F\'{\i}sica, Universidade Federal de
S\~ao Carlos, 13565-905 S\~ao Carlos SP, Brazil}

\date{\today}

\begin{abstract}
Polar optical phonons are studied in the framework of the
dielectric continuum approach for a prototypical
quantum-dot/quantum-well (QD/QW) heterostructure, including the
derivation of the electron-phonon interaction Hamiltonian and a
discussion of the effects of this interaction on the electronic
energy levels. The particular example of the CdS/HgS  QD/QW is
addressed and the system is modelled according to the spherical
geometry, considering a core sphere of material ``1" surrounded by
a spherically concentric layer of material ``2", while the whole
structure is embedded in a host matrix assumed as an infinite
dielectric medium. The strength of the electron-LO phonon coupling
is discussed in details and the polaronic corrections to both
ground state and excited state electron energy levels are
calculated. Interesting results concerning the dependence of
polaronic corrections with the QD/QW structure size are analyzed.
\end{abstract}

\maketitle

\section{Introduction}

Quantum-dot/quantum-well (QD/QW) semiconductor heterostructures
based on CdSe, ZnS, CdS, HgS, etc., have been synthesized by means
of colloidal solution chemistry in the latter years. They display
rather interesting size-dependent optical properties
\cite{b1,b1A,b2,b3,b3A,b3B} and appear to have promising
perspectives in the streaming progress of nanostructured device
technology. In the case of QD/QW heterostructures a few monolayers
of material ``2" (say, $\beta$-HgS) are epitaxially grown on a
core structure of material ``1" (say, CdS) and, after that, the
system is sometimes capped with an additional layer of material
``1". The whole system is embedded in a host material, which may
be a silicate matrix or an organic compound. We should remark
that, for QD/QW heterostructures based on  CdS/$\beta$-HgS, a true
quantum-well is created within the QD because the bulk $\beta$-HgS
energy gap is lower than the CdS gap. Highly confined carriers in
the $\beta$-HgS layer were actually detected by means of different
kinds of optical experiments. Besides the electronic and optical
properties, the vibrational spectra are rather important in the
discussion of the physical characteristics of these
heterostructures. Usually the so-called dielectric continuum
approach (DCA) has been applied for the theoretical determination
of TO and LO bulk-like phonons, while, in the case of small-size
crystal heterostructures, we must also consider the SO phonons
associated to the system interfaces \cite{b4}. Most of the
theoretical studies considered the spherical geometry
\cite{b4,b5,b6,b7,b8}, but in some cases spheroidal QD's were
investigated \cite{b9,b10,b10A}. Theoretically more reliable
approaches, as those reported in Refs.
[\onlinecite{b11,b11A,b11B}], involve rather complicated
calculations. However, it must be realized that the DCA, being an
essentially electrodynamical approach and relatively simple in its
theoretical formalism, is still able to reveal many important
physical features of the system as has been proved by all previous
works on the subject.

In this paper we focus on the confined internal LO phonons, and
also on the electron-LO phonon interactions for a QD/QW
heterostructure. The corresponding analysis for SO phonons was
already accomplished in Ref. [\onlinecite{b12}] for the same kind
of model-system. By applying the methods of the DCA we derive the
electric potential for confined internal LO phonons and, using
this result, the electron-LO phonon interaction Hamiltonian is
deduced. The electric potential allows to estimate the strength of
the interaction, but the interaction effects may also be estimated
by an analysis of the energy shifts of the unperturbed levels (the
so-called polaronic effect). We discuss that this kind of effects
is able to provide a better estimation of the electron-phonon
interaction strength. Polaronic effects have been studied by
different authors for systems more or less similar to the one we
are discussing in the present paper. Within the adiabatic
approximation, and using second order perturbation theory, the
polaronic corrections were studied for nanocrystallites in Refs.
[\onlinecite{b7,b12A,b12B}], and also in Refs.
[\onlinecite{b8,b12C,b12D,b12E}] by applying other methods. It is
worth to emphasize that nanostructured QD's, considering systems
beyond the single QD sphere, involve more complicated vibrational
properties and need to be further studied. Henceforth, in the
present work we aim to show the main effects of the electron-LO
phonon interactions and their influence on the energy levels of
the QD/QW heterostructure by applying a relatively simple model.
We show calculations involving both the ground state energy level
and the excited state energy levels of the QD/QW heterostructure
and comparisons are made with other previous works on the subject.

The paper is organized as follows. In Sec. II we give a brief
summary of the fundamental results of the DCA and apply them to
our model-system, thus deriving the internal LO phonon electric
potential. In Sec. III we analyze the wave functions and energy
levels in the framework of the effective mass approximation and
give a perturbative estimation of polaronic energy shifts.
Finally, Sec. IV presents a detailed discussion of the main
results of the work.

\section{Electric potential for the internal LO phonons}

The DCA is based on the well-known Born-Huang equations, which have been
extensively described in the literature \cite{b4,b5,b6,b9,b10} and leads
to the equation:
\begin{equation}
\varepsilon (\omega )\nabla ^{2}\varphi =0,  \label{e3}
\end{equation}
where $\varphi $ is the quasistatic electric potential and $\varepsilon
(\omega )$ is given by the standard expression: $\varepsilon (\omega
)=\varepsilon _{\infty }\left( \omega ^{2}-\omega _{L}^{2}\right) /\left(
\omega ^{2}-\omega _{T}^{2}\right) $. In the above equation the harmonic
time dependence $\varphi (\vec{r},\,t)\sim \exp (-i\omega t)$ is applied. We
further assume the Lyddane-Sachs-Teller relation: $\omega _{L}^{2}/\omega
_{T}^{2}=\varepsilon _{0}/\varepsilon _{\infty }$, while $\varepsilon _{0}$
 ($\varepsilon _{\infty }$) is the static (high frequency) dielectric
constant of the medium and $\omega _{L}$, $\omega _{T}$ are the
bulk longitudinal and transverse phonon frequencies at the
semiconductor $\Gamma $ point. The internal LO phonons satisfy Eq.
(\ref{e3}) with $\varepsilon (\omega ) = 0$ and, therefore,
$\omega=\omega_L$. Obviously, the electric potential does not
satisfy the Laplace equation. Thus, the potential must be obtained
as an expansion in an appropriate system of base functions
fulfilling the fundamental symmetry conditions of the structure.
In our case, the structure is spherically symmetric, and the
spherical coordinates ($r,\,\theta,\,\phi$) are the more
convenient. Of course, the electric potential must also satisfy
the usual boundary conditions: $\varphi $ should be continuous at
the interface between two different media and also should satisfy
the condition
\begin{equation}
\varepsilon _{1}\left[ \frac{\partial \varphi _{1}}{\partial n}\right]
_{S}=\varepsilon _{2}\left[ \frac{\partial \varphi _{2}}{\partial n}\right]
_{S},  \label{e4}
\end{equation}
i.e., the continuity of the normal component of the electric
displacement $\vec{D}$. Let us also recall the rather useful
relation
\begin{equation}
\nabla \varphi =\frac{\varepsilon _{\infty }\omega _{L}}{\varepsilon
_{\infty }-\varepsilon (\omega )}\sqrt{\frac{4\pi \rho }{\varepsilon ^{\ast }%
}}\vec{u},\quad \frac{1}{\varepsilon ^{\ast }}=\frac{1}{\varepsilon _{\infty
}}-\frac{1}{\varepsilon _{0}},  \label{e5}
\end{equation}
where $\rho $ is the reduced mass density addressed to the
involved ion couple and $\vec{u}$ is the mechanical relative
displacement vector with units of length. For the internal LO
phonons Eq.~(\ref{e5}) is applied taking $\varepsilon(\omega)=0$.

The geometry of the spherical QD/QW heterostructure is assumed in
the following form: in the interval  $0<r<a$ we have material ``1"
(a polar semiconductor like CdS), in the interval $a<r<b$ we have
material ``2" (a different polar semiconductor like HgS),
 and
finally for $r>b$ we shall assume an infinite dielectric medium
with a fixed dielectric constant $\varepsilon_D$. Using spherical
coordinates the electric potential is expanded as follows: for
$r<a$ we apply the complete set of orthogonal functions
$j_l(qr)Y_{lm}(\theta ,\,\phi )$, where $j_l(x)$ with
$l=0,1,2,\cdots$ are spherical Bessel functions and $Y_{lm}(\theta
,\,\phi )$ are the spherical harmonics with $|m|\leq l$; for
$a<r<b$ we shall apply the complete set of orthogonal functions
defined by $[n_l(qa)j_l(qr)-j_l(qa)n_l(qr)]Y_{lm}(\theta ,\,\phi
)$, where $n_l(x)$ are spherical Bessel functions of the second
kind. Let us remark that in the present model-system two different
semiconductor materials are involved and two different types of
phonons should be taken into account. In the central spherical
core ($r<a$) we have semiconductor ``1", oscillating with the
frequency $\omega_{1L}$, in the spherical layer ($a<r<b$) we have
semiconductor ``2" oscillating with the frequency $\omega_{2L}$.
For the same reason we must consider two different electric
potentials associated to these phonons: electric potential
$\varphi^{(1)}$, due to phonons in the core sphere, and electric
potential $\varphi^{(2)}$ due to phonons in the concentric
spherical layer. There is not overlapping between these electric
potentials: $\varphi^{(1)}$ is confined to the core sphere and
must be zero for $r \geq a$, while $\varphi^{(2)}$ is confined to
the spherical shell and must be zero for $r\leq a$ and $r\geq b$.
The latter properties of the electric potential are a clear
consequence of the boundary conditions (\ref{e4}) when applied to
the two spherical surfaces at $r=a$ and $r=b$ respectively. On the
other hand, the equation $j_l(qa)=0$ determines the possible
values of $q$ involved in the expansion of the electric potential
$\varphi^{(1)}$. For the electric potential $\varphi^{(2)}$ the
possible values of $q$ present in the corresponding expansion are
determined by the equation $n_l(qa)j_l(qb)=j_l(qa)n_l(qb)$.

It is useful to represent the electric potential in the following
form: $\varphi^{(i)} _{lm}(r,\,\theta ,\,\phi )=\Phi^{(i)}
_{l}(r)Y_{lm}(\theta ,\,\phi )$, where $\Phi_l^{(i)}(r)$
($i=1,\,2$) is the radial part of the potential, which is also
dependent on the various values of $q$ defining the possible
phonon modes. It is then clear that for $r<a$
\begin{eqnarray}
\Phi^{(1)}_l (r)&=&A^{(1)}_{lm}(q)j_l(qr), \label{e6}
\end{eqnarray}
and
\begin{eqnarray}
\Phi^{(2)}_l (r)&=&A^{(2)}_{lm}(q)\left
[n_l(qa)j_l(qr)-j_l(qa)n_l(qr)\right ], \label{e7}
\end{eqnarray}
for $a<r<b$. In Eqs. (\ref{e6}) and (\ref{e7}) $A^{(i)}(q)$ are
normalization constants (to be determined below) and the expansion
involves summations over $q,\,l,\,$ and $m$. In order to proceed
with the normalization of the electric potential we should
introduce the phonon annihilation and creation operators
$\hat{a}_{lm}(q)$ and $\hat{a}_{lm}^{\dagger }(q)$ obeying bosonic
commutation relations. Correspondingly, the classical displacement
vector $\vec{u}$ is transformed into the operator $\hat{\vec{u}}$
by means of
\begin{equation} \label{e8}
\hat{\vec{u}}^{(i)}=u^{(i)}_0\nabla f^{(i)}(\vec{r})\hat{a}_{lm}(q)
\end{equation}where
\begin{eqnarray}
f^{(1)}_{qlm}(\vec{r})&=&j_l(qr)Y_{lm}(\theta\,,\phi), \, \label{e9}\\
f^{(2)}_{qlm}(\vec{r})&=&\left [ n_l(qa)j_l(qr)-j_l(qa)n_l(qr)
\right ]Y_{lm}(\theta\,,\phi) \, . \label{e10}
\end{eqnarray}
We should realize that the $q$'s present in potentials of type
``1" or ``2" are different (determined by different transcendental
equations).

In order to determine the constants $u^{(i)}_{0}$ a well-known
procedure should be applied, which has been described elsewhere
\cite{b5,b9,b10,b12}. We transform the classical kinetic energy
due to the vibrations into a quantum operator, which is then taken
as the LO phonon Hamiltonian $\hat{H}_{ph}$. Requiring this
Hamiltonian to have its standard harmonic form, i.e.,
$\hat{H}^{(i)}_{ph}=\hbar
\omega_{iL}(\hat{a}^{\dagger}\hat{a}+1/2)$ we can determine
$u^{(i)}_0$. The calculations are straightforward and we briefly
show them in the Appendix. We have found that

\begin{eqnarray}
u^{(1)}_0&=& \left [ \frac{4\hbar }{\rho _{1}\omega_{1L}ax_1^2j_{l+1}^2(x_1)}\right ]^{1/2}\, ,  \label{e11}\\
u^{(2)}_0&=& \left [ \frac{4\hbar }{\rho _{2}\omega_{2L}ax_2^2M_l(x_2)}\right ]^{1/2}\, ,  \label{e12}
\end{eqnarray}
where
\begin{eqnarray}
M_{l}(x) &=&\gamma ^{3}\left[ j_{l}(x)n_{l+1}(\gamma
x)-n_{l}(x)j_{l+1}(\gamma x)\right] ^{2}  \nonumber \\
&&-\left[ j_{l}(x)n_{l+1}(x)-n_{l}(x)j_{l+1}(x)\right] ^{2}\,,\,
\label{e13}
\end{eqnarray}%
and $x=qa$ are the roots of the corresponding transcendental
equation: $j_l(x)=0$ for phonons of type ``1" and
$n_l(x)j_l(\gamma x)=j_l(x)n_l(\gamma x)$ for phonons of type
``2". We have also introduced the parameter $\gamma = b/a$.

By the above described method we finally obtain the normalization
constants $A^{(i)}_{lm}(q)$ (notice that  actually they do not
depend on $m$ due to the azimuthal symmetry of our model)
introduced in Eqs. (\ref{e6}) and (\ref{e7}) . In this step Eq.
(\ref{e5}) must be applied and we obtain
\begin{eqnarray}
A^{(1)}_{l}(x_1)&=&\left [\frac{16 \pi \hbar \omega_{1L}}{\varepsilon^{*} _{1}ax_1^2j_{l+1}(x_1)^2}\right] ^{1/2}\, ,  \label{e14}\\
A^{(2)}_{l}(x_2)&=&\left [\frac{16 \pi \hbar \omega_{2L}}{\varepsilon^{*} _{2}ax_2^2M_l(x_2)^2}\right] ^{1/2}\, .  \label{e15}
\end{eqnarray}

In Eqs. (\ref{e14}) and (\ref{e15})
$1/\varepsilon^{*}_i=1/\varepsilon_{i\infty}-1/\varepsilon_{i 0}$.
We have thus obtained an explicit analytic expression for the
electric potential operator $\hat{\varphi}^{(i)}$ corresponding to
both types of internal LO phonons involved in our model-system:
\begin{eqnarray}
\hat{\varphi}^{(1)}_{qlm}(r,\,\theta ,\,\phi )&=&\frac{1}{2}A^{(1)}_{l}(x_1)\left [ f^{(1)}_{qlm}(\vec{r})\hat{a}_{lm}(x_1)+ h.c.\right ] \label{e16}\\
\hat{\varphi}^{(2)}_{qlm}(r,\,\theta ,\,\phi
)&=&\frac{1}{2}A^{(2)}_{l}(x_2)\left [
f^{(2)}_{qlm}(\vec{r})\hat{a}_{lm}(x_2)+ h.c.\right ] \label{e17}
\end{eqnarray}
where `h.c.'$\,$ stands for `hermitian Conjugate'. Let us once
again emphasize that the electric potential operator with $i=1$ is
defined as the zero operator for $r>a$, while the operator with
$i=2$ is the zero operator for $r<a$ and $r>b$. The electric
potential operators defined in Eqs.(\ref{e16}) and (\ref{e17}) are
hermitian and allow us to define the electron-LO phonon
interaction Hamiltonians, which are given by
\begin{equation}
\hat{H}^{(i)}_{e-ph}=-e\sum_{qlm}\hat{\varphi}^{(i)}_{qlm}(r,\,\theta ,\,\phi )\, ,
\label{e18}
\end{equation}
where $e$ is the absolute value of the electron charge. In
Eq.(\ref{e18}) we have included a summation over all the possible
eigenfunctions of this model-problem. In some expressions we are
using simultaneously the notations $q$ or $x=qa$ in order to
denote the roots of the corresponding transcendental equations.
Whenever the notation $x_i$ is applied the roots of the
corresponding modes for $i=1,\,2$ are implied.  This should not
lead to confusions.

\section{Electron energy levels and polaronic corrections}

Applying the effective mass approximation for the electronic
states, we model the confining potential for the electron as

\be \label{e19}U(r)=\left \{ \begin{array} {llll}
0 &\, , \, \mbox{for} & 0< r <a \\
-U_0&\, , \, \mbox{for} & a<r<b \\
\infty &\, , \, \mbox{for} & r \geq b
\end{array} \right.
\ee where $U_0$ is the band-offset between the conduction bands of
semiconductors ``1" and ``2". We are here assuming that
semiconductor ``2" has a smaller energy gap and, therefore, in the
shell layer $a<r<b$ the conduction electrons are actually
subjected to a QW. The zero energy level was chosen at the bottom
of the semiconductor ``1" conduction band.

The solution of the Schr\"odinger equation
$\hat{H}\Psi(\vec{r})=E\Psi(\vec{r})$ is a straightforward problem
and we are led to the wave function

\be \label{e20}
\Psi(r,\,\theta,\,\phi)=\psi(r)Y_{lm}(\theta,\,\phi)\, , \ee where
\be \label{e21} \psi(r)=C\left \{ \begin{array} {lll}
j_l(k_0r) & , \,&\mbox{for}\, \, r \leq a \\
j_l(k_0a) \left
[\frac{n_l(kb)j_l(kr)-j_l(kb)n_l(kr)}{n_l(kb)j_l(ka)-j_l(kb)n_l(ka)}
\right ] & , \, &\mbox{for}\, \,a\leq r \leq b
\end{array} \right.
\ee In the above equation we have defined
\begin{eqnarray}
k_0 &=& \sqrt{\frac{2\mu_1E}{\hbar^2}}\, , \label{e22}\\
 k&=& \sqrt{\frac{2\mu_2(U_0+E)}{\hbar^2}}\, , \label{e23}
\end{eqnarray}
where $\mu_i$ with $i=1,\,2$ denote the electron effective mass
for each semiconductor. In this context the symbol $\mu_0$ stands
for the free electron mass. The energy levels, denoted by $E$, are
all discrete and could be positive or negative. For negative
energy levels (always larger than $-U_0$) we should take
$k_0=i\kappa$, where $\kappa$ is again given by (\ref{e22}) taking
$|E|$ in the place of $E$. The normalization constant $C$ can be
calculated requiring $\int |\Psi|^2d^3r=1$.

Notice that the wave-function described in (\ref{e21}) is
continuous at $r=a$ and equal to zero at $r=b$. The additional
boundary condition

\be \label{e24} \frac{1}{\mu_1}\left [\frac{\partial
\psi_1}{\partial r}\right ]_a=\frac{1}{\mu_2}\left [\frac{\partial
\psi_2}{\partial r}\right ]_a \, . \ee leads to the discrete
energy levels, given by the roots of the transcendental equation

\be \label{e25}
\frac{k_0\mu_2}{k\mu_1}\frac{j_l\p(k_0a)}{j_l(k_0a)}=\frac{n_l(kb)j_l\p(ka)-j_l(kb)n_l\p(ka)}{n_l(kb)j_l(ka)-j_l(kb)n_l(ka)}\,
. \ee
In Eq.(\ref{e25}) the prime in the spherical Bessel
functions denotes the first derivative with respect to the
function's argument.

It is interesting to analyze the effects of the electron-LO phonon
interaction on the electron energy levels. For the considered
semiconductors this interaction is, in general, weak and should
lead to a relatively small shift towards the lower energies.
Taking into account just the interaction with confined internal LO
phonons we realize that, for the studied QD/QW heterostructure,
two different electron-LO phonon interaction Hamiltonians are
involved, as predicted by Eq.(\ref{e18}). For LO phonons of type
``1" (``2") only the region $r<a$ ($a<r<b$) provides a
contribution. In general, the SO phonons (analyzed in Ref.
[\onlinecite{b12}]) also contribute to the energy shift for all
the electronic states with $l\geq 1$. For electron states with
$l=0$ the SO phonons do not contribute to the energy shift. Other
important issue is the degeneracy of electron states with $l\geq
1$. This degeneracy (of order $2l+1$) in general could be lifted
if the interaction with LO and SO phonons is taken into account.
In the present work spin effects are not considered.

The general theory for the calculation of the polaronic
corrections in systems similar to the one studied in this work has
been described elsewhere \cite{b8,b12A,b12B,b12C,b12D} and need
not be given here in detailed form. Applying the perturbative
approach (and assuming that the system behaves adiabatically) we
are led to the expression

\be \label{e26} \Delta E_p=-\sum_{p\p}\sum_i\frac{\left | \left
\langle p\p , \, 1 | \hat{H}^{(i)}_{e-ph}|p, \, 0 \right \rangle
\right |^2}{E^0_{p\p}+\hbar\omega_{iL}-E^0_{p}}. \ee In
Eq.(\ref{e26}) we are assuming $T=0$ K, so all the LO phonons
involved in the polaronic corrections are virtual. Label ``$p$" is
used to describe the electronic state we are analyzing, while
label ``$p\p$" denotes all the intermediate electronic states in
the second order perturbation theory formula. $\Delta E_p$ gives
the energy shift of level ``$p$" due to the electron-phonon
coupling. The symbol $E^0_p$ corresponds to unperturbed electron
energies and $| p,\, 1\rangle$ ($| p, \, 0 \rangle $) is a
ket-vector notation for the electron-one phonon (electron-zero
phonon) states. In our concrete calculations the symbol ``$p$"
involves three quantum numbers ($l,\,m,\,\nu$) but the energy
levels depend just on ``$l,\,\nu$", where ``$\nu$" denotes the
roots of the corresponding transcendental equation
(Eq.(\ref{e25})) for a fixed value of $l$. If we are renormalizing
the energy of an electron state with $l=0$ the calculations become
easier because the corresponding level is not degenerate and the
contribution from SO phonons can be ignored. States with $l\geq 1$
are obviously more complicated and require quite involved
calculations. Another important issue present in Eq.(\ref{e26}) is
the summation over all possible phonon states. However, it should
be kept in mind that the possible phonon modes are, in general,
bounded, i.e. the number of possible modes cannot be larger than
the system volume divided by the unit cell volume. In the
practical case, just the first phonon modes provide a significant
contribution to the energy shift.

\section{Discussion of the results}

Let us discuss the essential results obtained in the foregoing
sections. For the numerical computations we considered that
material ``1" is CdS and material ``2" is $\beta$-HgS. The values
of the corresponding physical parameters are given in Table~I.
\vspace{3mm}

\noindent {\bf Table I}: Semiconductor physical parameters.
\vspace{-2mm}
\begin{center}
\begin{tabular}{|c|c|c|c|c|c|c|}
\hline  & $\varepsilon_{\infty}$ & $\varepsilon_0$ & $\omega
_{T}($cm$^{-1})$ & $\omega _{L}($cm$^{-1})$ & $\mu / \mu_0$ &
$E_g ($eV$)$ \\ \hline CdS & 5.5 & 9.1 & 233 & 300 & 0.2 & 2.5 \\
\hline HgS & 11.36 & 18.2 & 197.5 & 250 & 0.036 & 0.5 \\ \hline
\end{tabular}
\end{center}

We must emphasize that, within the adopted model, the properties
of the dielectric host medium for $r>b$ are immaterial. Both
internal LO phonons and electrons are defined for $r<b$. SO
phonons, however, do depend on the properties of the host medium,
but they are not considered here.

\begin{figure}[tbp]
\includegraphics*[width=0.8\linewidth]{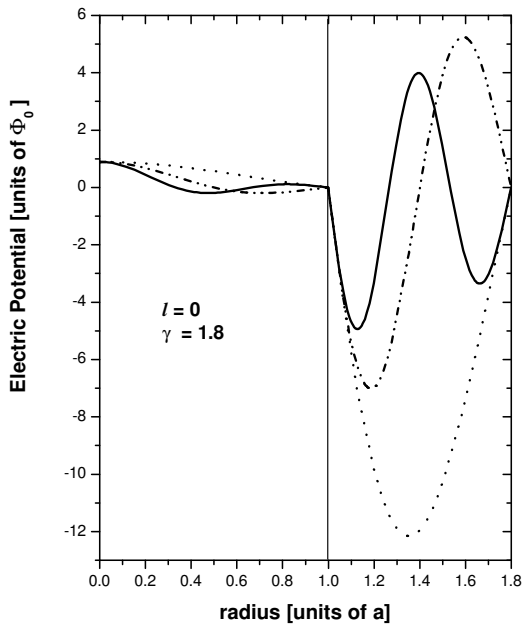}
\caption{Radial part of the electric potential (in units of
$\Phi_0$) as a function of the radius (in units of $a$) for $l=0$
and $\gamma = 1.8$. The potentials correspond to the first three
LO phonon modes and are depicted by a dotted, dash-dotted and
solid line respectively.} \label{1}
\end{figure}

\begin{figure}[tbp]
\includegraphics*[width=0.8\linewidth]{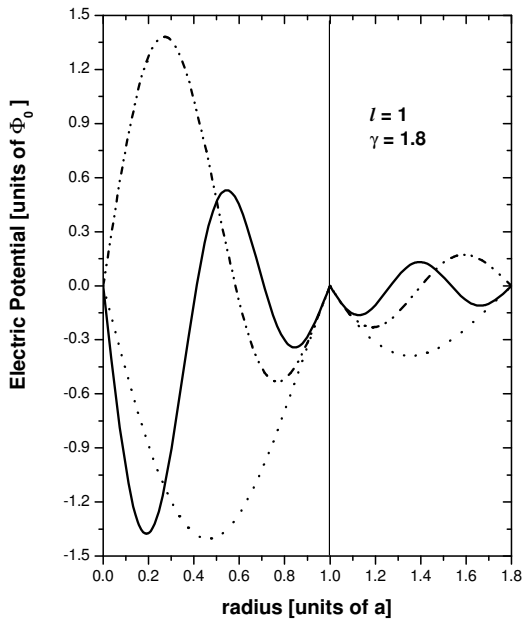}
\caption{Same as in Fig.~1 but now displaying the potentials for
the first three LO phonon modes with $l=1$. } \label{2}
\end{figure}

One important issue which deserves a close analysis is the
strength of the electric potential associated to the two different
types of internal LO phonons of our system. A first estimation can
be made by a study of the radial part of the electric potentials
as given in Eqs. (\ref{e6}) and (\ref{e7}). In Fig.~1 we are
plotting $\Phi^{(i)}(r)$ (in units of
$\Phi_0=(4\hbar\omega_{1L}/\pi a\varepsilon_1^{*})^{1/2}$) for
$l=0$, setting $\gamma = 1.8$ and taking into account the first
three roots of the corresponding transcendental equations:
$j_l(x)=0$ or $n_l(x)j_l(\gamma x)=n_l(\gamma x)j_l(x)$ for $i=1$
or $2$. In the figure we can see the dependence of the radial part
of the potential on $r$, and two regions are displayed. For the
first part ($r<a$) what we actually have is the electric potential
$\Phi^{(1)}(r)$, due to LO phonons of type ``1". In the second
part ($a<r<b$) the displayed potential is $\Phi^{(2)}(r)$, due to
LO phonons of type ``2". In Fig.~1 we have used a dotted, a
dash-dotted, and a solid line in order to distinguish the three
first roots of the transcendental equations in increasing order.
As expected, for higher values of the corresponding root the
electric potential oscillates more intensively, an effect that
should led to lower values of the electron-phonon matrix elements.
On the other hand, it is interesting to note that the intensity of
the potential, i.e., the absolute value of $\Phi^{(i)}(r)$,  is
lower within the core sphere ($r<a$) in comparison with the layer
region $a<r<b$. When we limit ourselves to a discussion of just
the radial part of the potential, we are loosing some information
concerning the angular dependence that is actually non trivial.
However, the radial part provides a certain estimation of the
potential strength that gives an insight into the fundamental
characteristics of the electron-LO phonon interaction. In Fig.~2
we analyze the radial part of the potential taking now $l=1$ and
the same value for $\gamma$. Again the first three roots of the
transcendental equations were considered for this value of $l$. It
is evident from the figure that, in contrast with the results
shown in Fig.~1, in the present case the potential intensity is
higher inside the core region. From the latter results it could be
inferred that the electron should feel a stronger phonon
interaction in the region $r<a$ for $l=1$, while for $l=0$ the
electron-phonon interaction should be stronger in the region
$a<r<b$. However, this kind of analysis could be certainly
misleading. In order to properly understand the strength of the
electron-phonon interaction we must also take into account the
electron wave functions. For instance, if for $l=0$ we get a
larger probability amplitude of finding the electron inside the
core sphere, then this region could provide the larger
contribution to the electron-phonon interaction. For that reason
the strength of the interaction is better revealed in such
quantities as relaxation rates, polaronic corrections, etc., where
the matrix elements of the electron-phonon interaction Hamiltonian
must be calculated.

\begin{figure}[tbp]
\includegraphics*[width=0.8\linewidth]{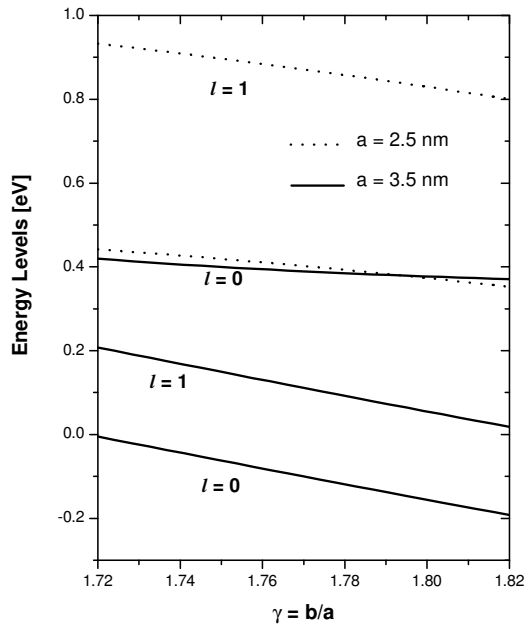}
\caption{Energy levels for the unperturbed CdS/HgS QD/QW
heterostructure (in eV) as a function of $\gamma = b/a$. For
$a=2.5$ nm we show the first two levels (dotted lines). For
$a=3.5$ nm (solid lines) the first three energy levels are shown.
For each level the corresponding value of $l$ is explicitly
indicated.} \label{3}
\end{figure}

In Fig.~3 we are showing (solid lines) the first three energy
levels of the QD/QW heterostructure as a function of $\gamma =
b/a$. These results were obtained from Eq.(\ref{e25}) without the
inclusion of the electron-phonon (polaronic) corrections. For the
core sphere radius we have taken $a=3.5$ nm, while the band-offset
between the conduction bands of CdS and $\beta$ -HgS was estimated
as $U_0=1.34$ eV following Ref. [\onlinecite{b13}]. For this value
of $a$ the first (ground state) energy level is localized inside
the QW region ($E<0$ for our choice of the zero energy level at
the bottom of the CdS conduction band). The first excited energy
level (with $l=1$) is already out of the QW with $E>0$ and the
second excited level (with $l=0$) can also be seen in the figure.
The values of $\gamma$ were chosen in a relatively narrow interval
$1.72<\gamma <1.82$. According to reported experimental results
\cite{b1,b1A,b2,b3}, $\gamma$ is bound to be not very large. On
the other hand, for lower values of $\gamma$ there are not
possible levels for $E<0$. For this energy level the wave function
in the region $r<a$ is expressed through hyperbolic functions,
while for other energy levels wavefunctions are given by
oscillating trigonometric functions. The latter fact provides
important changes in the calculation of matrix elements and leads
to differences in polaronic energy shifts (to be discussed below).
In Fig.~3, using dotted lines, we show the first two energy levels
for a core sphere radius with $a=2.5$ nm. It is clear from the
figure that, in this latter case, there are not energy levels with
$E<0$ in the analyzed interval for $\gamma$. In Fig.~3 we are not
showing the third energy level for $a=2.5$ nm with $l=0$. This
level has the energy $E\approx 1.30$ eV when $\gamma = 1.72$
decreasing down to $E\approx 1.08$ eV for $\gamma =1.8$.

As discussed in Section III, the electron-phonon interaction leads
to an energy shift of the electronic levels towards the lower
energies, the so-called ``polaronic effect". Polaronic effects in
the electron, hole, and exciton energy levels for QD structures
have been studied by different authors
\cite{b7,b8,b12A,b12B,b12C,b12D}. In most of the published papers
the single spherical QD was considered, while the applied
theoretical formalism is in some cases perturbative and in other
cases the variational one. For a system like the one studied in
the current paper, it is plausible to apply a perturbative
approach as was briefly described in Section III. Polaronic
corrections may be discussed considering both the ground state
energy level and the excited states levels, and usually involve
tedious lengthy calculations. In order to make our calculations we
applied Eq.(\ref{e26}) where the labels $p$ and $p\p$ were adapted
to our system. The electron-LO phonon interaction was evaluated
according to Eq.(\ref{e18}) considering both types of LO phonons
present in the QD/QW heterostructure and including summations over
an appropriate amount of phonon modes. On the other hand, electron
wavefunctions, properly normalized, were taken according to
Eq.(\ref{e21}) and the discrete energy levels were calculated by
solving the transcendental equation (\ref{e25}).

\begin{figure}[tbp]
\includegraphics*[width=0.8\linewidth]{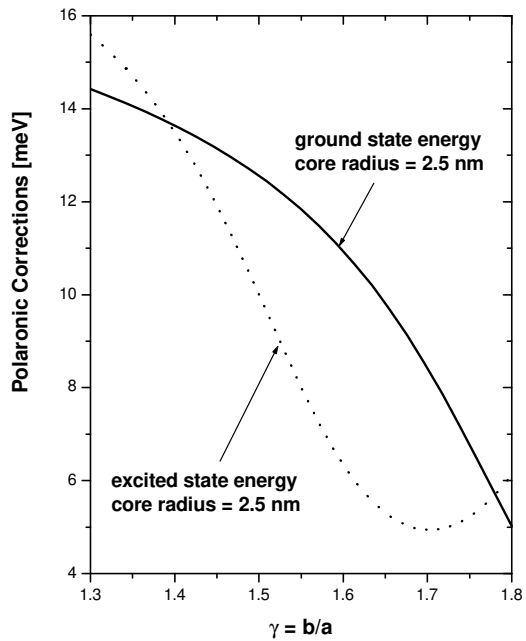}
\caption {Absolute value of the energy shifts (in meV) as a
function of $\gamma = b/a$ for the polaronic corrections to the
ground state (solid line) and excited state (dotted line)  energy
levels considering a QD/QW heterostructure with core radius
$a=2.5$ nm; These energy shifts should be subtracted from the
corresponding energy levels presented in Fig.~3.} \label{4}
\end{figure}

In Fig.~4 we present the absolute value of the polaronic energy
shift corresponding to the case with $a=2.5$ nm. We show the
corrections to the ground state energy level (the first dotted
curve in Fig.~3) for a relatively wide range of $\gamma$ values.
These energy shifts, given here in meV, should be subtracted from
the corresponding energies reported in Fig.~3. The energy shifts
are decreasing for growing values of $\gamma$, while the
calculated values are relatively low as could be expected for a
system having a weak electron-LO phonon coupling. The obtained
results disagree with those reported in Ref. [\onlinecite{b12C}]
for a similar situation in which the opposite trend for the
polaronic corrections was obtained. In Fig.~4 we also show the
polaronic corrections to the excited energy level (dotted curve).
This energy level, as mentioned above, has $l=0$ and was not shown
in Fig.~3. The curve, giving the dependence with $\gamma$, shows a
decreasing behavior in its first part but, after reaching a
minimum approximately at $\gamma = 1.7$, shows an increasing
behavior for the last part of the considered interval. This
peculiar kind of dependence with the thickness of the HgS layer
seems to characterize the polaronic corrections for the excited
levels and was in a certain way commented in Ref.
[\onlinecite{b12D}]. It must also be remarked that, for most of
the $\gamma$ interval, the corrections to the excited level are
lower, than those to the ground state energy level.

In Fig.~5 we present the same dependencies as in Fig.~4, but
taking now $a=3.5$. As easily seen in the figure the polaronic
corrections to the excited state (considering again the state with
$l=0$, i.e., the third solid line in Fig.~3) are now considerably
higher than the corrections related to the ground state energy
level. Actually, the polaronic energy shifts corresponding to the
ground state energy level are quite low and difficult to detect in
the experimental measurements. The latter results, in sharp
contrast with those described in Fig.~4, can be easily understood
from the physical point of view. The ground state wave function
(for $a=3.5$ nm) in the region $r<a$ is given by hyperbolic
functions (in the framework of the used model-system), thus
leading to very low probability amplitudes if compared with states
having $E>0$. We must also notice that in Fig.~4 we are
considering a narrower interval for $\gamma$ in order to ensure
the presence of the ground state energy with $E<0$.

\begin{figure}[tbp]
\includegraphics*[width=0.8\linewidth]{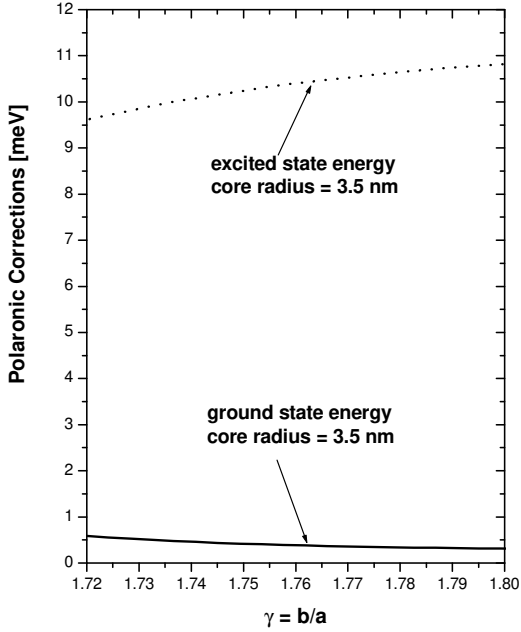}
\caption{Same as in Fig.~4 but taking $a=3.5$ nm. In both cases
the results correspond to the CdS/HgS QD/QW heterostructure.}
\label{5}
\end{figure}

As general conclusions to the present work, we can emphasize that
polaronic corrections are actually interesting in the study of the
electronic energy spectra of QD/QW heterostructures. They can be
detected in the appropriate optical experiments, differences
between the ground state and excited states polaronic corrections
may be studied and should provide good estimations of the
electron-LO phonon coupling in this kind of heterostructures.
Concerning degenerate energy levels (not studied in the current
work) the electron-LO phonon interaction could introduce
detectable energy splittings between the degenerate levels, a
subject that requires further analysis.

\appendix{Normalization of the electric potential}

When we substitute the hermitized operator version of $\vec{u}^2$,
i.e., $\vec{u}^2\to
(\hat{\vec{u}}^{\dagger}\cdot\hat{\vec{u}}+\hat{\vec{u}}\cdot\hat{\vec{u}}^{\dagger})/2$,
in the classical expression for the kinetic energy due to the ion
vibrations, we are directly led to: \be \label{A1}
\hat{H}^{(i)}_{ph}=\frac{1}{2}\rho_i\omega^2_{iL}u^2_{i0}\int_{V_i}\nabla
f^{(i)*}\cdot \nabla f^{(i)} \,d^3r
\,(\hat{a}^{\dagger}\hat{a}+1/2)\, , \ee where Eq.(\ref{e8}) was
applied and $f^{(i)}(\vec{r})$ are the two functions given in Eqs.
(\ref{e9}) and (\ref{e10}). In the above expressions, for the sake
of brevity, we are avoiding some subscripts. The volumes $V_i$
are: for $i=1$ the sphere of radius $a$ and for $i=2$ the
spherical shell defined by $a<r<b$.  Integrating by parts the
integral in Eq.(\ref{A1}), we first obtain a surface integral that
is always zero because of the boundary conditions fulfilled by the
mentioned functions (notice that these functions are zero over the
spherical surfaces defining the boundaries of volumes $V_1$ and
$V_2$ respectively). Taking into account that \be \label{A2}
\nabla^2 f^{(i)}= -q^2 f^{(i)}\, , \ee the integral can be casted
as
 \be \label{A3} \int_{V_i}\nabla f^{(i)*}\cdot \nabla f^{(i)}
\,d^3r=q^2\int_{V_i}f^{(i)*}f^{(i)}d^3r \, . \ee The latter
integral contains an integration over the angular part of the
functions that is easily evaluated in the form: $\int
Y^{*}_{lm}Y_{lm}d\Omega=1$. There remain two integrals over the
radial parts that are different for $i=1,\,2$ and given by

\be \label{A4} \int_0^a j_l^2(qr)r^2dr =
\frac{a^3}{2}j^2_{l+1}(qa), \ee \be \label{A5} \int_a^b\left[
n_l(qa)j_l(qr)-j_l(qa)n_l(qr)\right]^2r^2dr = \frac{a^3}{2}M_l(qa)
\end{equation}
where $M_l(x)$ was given in Eq.(\ref{e13}). For the integral
(\ref{A4}) the condition $j_l(qa)=0$ is assumed, while for the
integral (\ref{A5}) the corresponding condition is
$n_l(qa)j_l(qb)=j_l(qa)n_l(qb)$. The latter results, together with
the analysis made in Section II, lead to the values of
$u^{(i)}_{0}$ (reported in Eqs. (\ref{e11}) and (\ref{e12})).
Using now the relation
\be \label{A6} A^{(i)}_l(q)=\sqrt{4\pi
\rho_i\omega_{iL}/\varepsilon^{*}_i}u_0^{(i)} \ee we finally
obtain the normalization constants for the electric potential
reported in Eqs. (\ref{e14}) and (\ref{e15}).

\begin{acknowledgments} The work is partially supported by grants from the
FAPESP (Funda\c{c}\~ao de Amparo \`a Pesquisa de S\~ao Paulo) and
Conselho Nacional de Desenvolvimento Cient\'{i}fico e
Tecnol\'{o}gico (CNPq). F.C. is grateful to Departamento de
F\'{\i}sica, Universidade Federal de S\~ao Carlos, for
hospitality.
\end{acknowledgments}

\end{document}